\documentclass[a4paper,11pt]{article}
\pdfoutput=1 

\usepackage{jheppub} 
\usepackage{multirow}
\usepackage[T1]{fontenc} 
\usepackage{color}
\usepackage{lineno}

\title{\boldmath\huge{Sterile neutrino oscillometry with Jinping} }

\author[a,1]{M.V. Smirnov,\note{Corresponding authors}}
\author[a,1]{Zh.J. Hu,}
\author[a,b,1]{J.J. Ling,}
\author[c,d]{Yu.N. Novikov,}
\author[e]{Z. Wang}
\author[f,1]{{\rm and} G. Yang}

\affiliation[a]{School of physics, Sun Yat-Sen University, Guangzhou 510275, China}
\affiliation[b]{Key Laboratory of Particle \& Radiation Imaging (Tshinghua University), \\Ministry of Education, Beijing 10084, China}
\affiliation[c]{Petersburg Nuclear Physics Institute, Gatchina, St. Petersburg 188300, Russia}
\affiliation[d]{Saint-Petersburg State University, Peterhof, St.Petersburg 198504, Russia}
\affiliation[e]{Department of Engineering Physics, Tsinghua University, Beijing 100084, China}
\affiliation[f]{Department of Physics and Astronomy, State University of New York at Stony Brook, \\Stony Brook, New York 11794, USA}

\emailAdd{gear8mike@gmail.com}
\emailAdd{huzhj3@mail2.sysu.edu.cn}
\emailAdd{lingjj5@mail.sysu.edu.cn}
\emailAdd{guang.yang.1@stonybrook.edu}

\abstract
{
The existence of sterile neutrino is an open question in neutrino physics up to now.
The method of neutrino oscillometry provides a powerful tool to test the common 3+1 sterile neutrino hypothesis, i.e. three active flavors and one sterile falvor.
There are several antineutrino sources can be used for this method.
One of them is the well known isotope chain of $^{144}{\rm Ce}-$$^{144}{\rm Pr}$ with initial activity around 50-100 kCi.
It has compact size and might be installed either outside or inside the detector.
Another one is the short-lived isotope $\rm^8Li$, which can be produced in nuclear reaction of a proton beam hitting beryllium target. 
The Lithium source has only the out-of-detector option due to its large size.
The proposed Jinping water-based liquid scintillator detector will be used as a detection volume.
Above experimental setups will allow us to cover the current best fit values of oscillation parameters with 90\% C.L.
At the same time, it is sensitive to the region of Neutrino-4 result.
}

\begin{document}	
\maketitle
\flushbottom

\section{Introduction}
\label{sec:intro}

\par
\indent

It is known that neutrino is the second most prevalent particle in the Universe.
Significant progress has been achieved for neutrino physics in the last two decades.
Successful observations of neutrino oscillations have confirmed that neutrinos are massive particles~\cite{sup_kam}.
The majority of experimental results are in good agreement with the theory of three-neutrino oscillations.
However several experiments and their results cannot be explained up to now.
These results are called ``anomalies''.
The first so-called ``gallium anomaly'' was observed in solar neutrino experiments GALLEX and SAGE~\cite{gallex,sage}.
The second anomaly result came from two beam experiments LSND and MiniBooNE~\cite{lsnd,miniB}.
The latest result from MiniBooNE has increased the tension between oscillation theory and experimental data~\cite{Aguilar-Arevalo:2018gpe}.
Recently, another ananomly has appeared after revisions of the reactor data~\cite{re_anom}.
One possible explanation for these anomalies might be hidden or uncounted uncertainties inside the experimental setup.
Another possible solution is introducing one or several sterile flavors, which can mix with the standard active flavors.
This is assuming that the sterile neutrino has a mass in the eV-scale.
It should be noted that from the decay width of Z-boson only three active flavors of neutrino may exist~\cite{PDG}.
There are several possible schemes of mixing between sterile and active flavors of neutrino.
The simplest scenario of mixing is the so called 3+1 scheme, where three active flavors of neutrino and one sterile state are involved.
For this case it is reasonable to use the short baseline limit, when leading contribution to oscillations comes only from sterile oscillation parameters (mixing angle and mass square difference).
Other oscillation parameters will not impact the oscillation probability.
For the 3+1 scheme, the survival oscillation probability of electron neutrino (anti-neutrino) can be written in the following form~\cite{Abazajian:2012ys}:
\begin{equation}
\label{eq_1}
P(\nu_e\rightarrow\nu_e) \approx 1- \sin^2(2\theta_{14})\cdot\sin^2\Big(1.27\cdot\Delta m^2_{41}\cdot\frac{L[{\rm m}]}{E[{\rm MeV}]}\Big),
\end{equation}
where $\theta_{14}$ is new neutrino mixing angle; $\Delta m^2_{41}$ is the mass difference between the fourth and first neutrino mass states.
Thus from Eq.~\eqref{eq_1} it follows that for eV-scale neutrinos the presence of sterile neutrinos can be detected as a deficit in total event rate and distortion in neutrino spectrum shape. 
As the oscillation frequency is relatively high in comparison with ordinary oscillations, short baseline experiments with high-intensity neutrino flux are required in searching for the sterile neutrino.
All of the above conditions are fulfilled by the method of neutrino oscillometry.
This method is described below.

\section{The method of neutrino oscillometry}
\label{sec_2}

\par
\indent

The core of this method is based on the concept of oscillation length and the neutrino detection inside the detector fiducial volume.
Including the sterile neutrino, the oscillation length can be expressed as~\cite{Novikov:2011gp}:
\begin{equation}
\label{eq_2}
L_{\rm osc}=\frac{\pi E[{\rm MeV}]}{1.27\cdot\Delta m^2_{41}}.
\end{equation}
Here oscillation length is the distance between two adjacent highs.
Hence if the detector size is smaller than $L_{\rm osc}$, the direct observation of the oscillation curve will be nearly impossible for the assumed experimental setup.
As the neutrino is a weak interacting particle, the cross-section of its interaction with detector material is tiny.
Therefore, the incoming neutrino flux should be well known and intensive enough to provide the shape measurements of the neutrino spectrum.
Reactor or beam neutrinos (antineutrinos) are not suitable for such kinds of measurements.
However a high-intensity handmade artificial source can be used for oscillometry purposes.
Originally this idea was proposed for $\theta_{13}$ measurement using the monoenergetic neutrino emitters and later for sterile neutrino searching~\cite{Novikov:2011gp}-\cite{Smirnov:2015rha}.
The main candidate for the neutrino source is a well known isotope $^{51}{\rm Cr}$, which has a few disadvantages: required huge activity around 5 MCi in order to get rid of the overlapping solar background, intense gamma radiation and a short life-time 27.7 days~\cite{gallex,sage}.
Later, antinetrino emitter was investigated~\cite{Ce_source}.
Isotope $^{144}{\rm Ce}-$$^{144}{\rm Pr}$ was proposed as a possible source for the sterile neutrino experiments.
SOX experiment from Borexino was aimed to implement this idea, but unfortunately it had a hard time of the source production and then the project has been canceled~\cite{Borexino:2013xxa}.
Hereinafter we focus on $^{144}{\rm Ce}-$$^{144}{\rm Pr}$ as an antineutrino source for the future solar neutrino experiment Jinping~\cite{Jinping}. Another elegant suggestion is to use a short-lived isotope of $\rm^8Li$ as an intensive antineutrino emitter~\cite{lithium_source}.
The main proposal about this source is called IsoDAR and has been proposed for the KamLAND detector~\cite{Abs:2015tbh}.
Besides, the method of neutrino oscillometry might provide a direct test in case of existence more than one sterile neutrino flavor.

\subsection{Antineutrino sources}
\par
\indent

\subsubsection{$^{144}{\rm Ce}-$$^{144}{\rm Pr}$ source}
\begin{figure}[ht]
\centering
\begin{minipage}[b]{0.44\linewidth}
\centering
\includegraphics[scale=0.5]{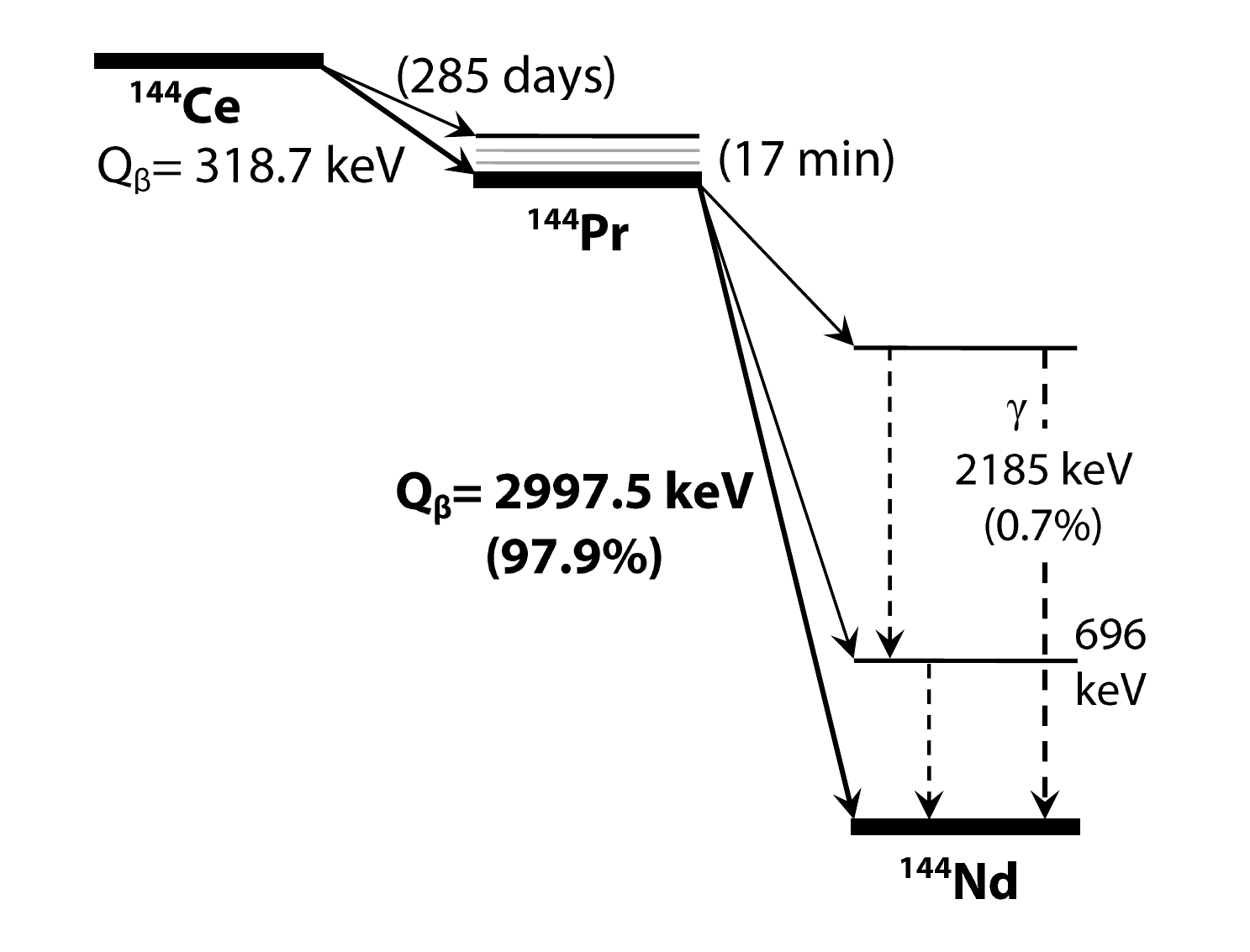}
\end{minipage}
\qquad
\begin{minipage}[b]{0.48\linewidth}
\centering
\includegraphics[scale=0.4]{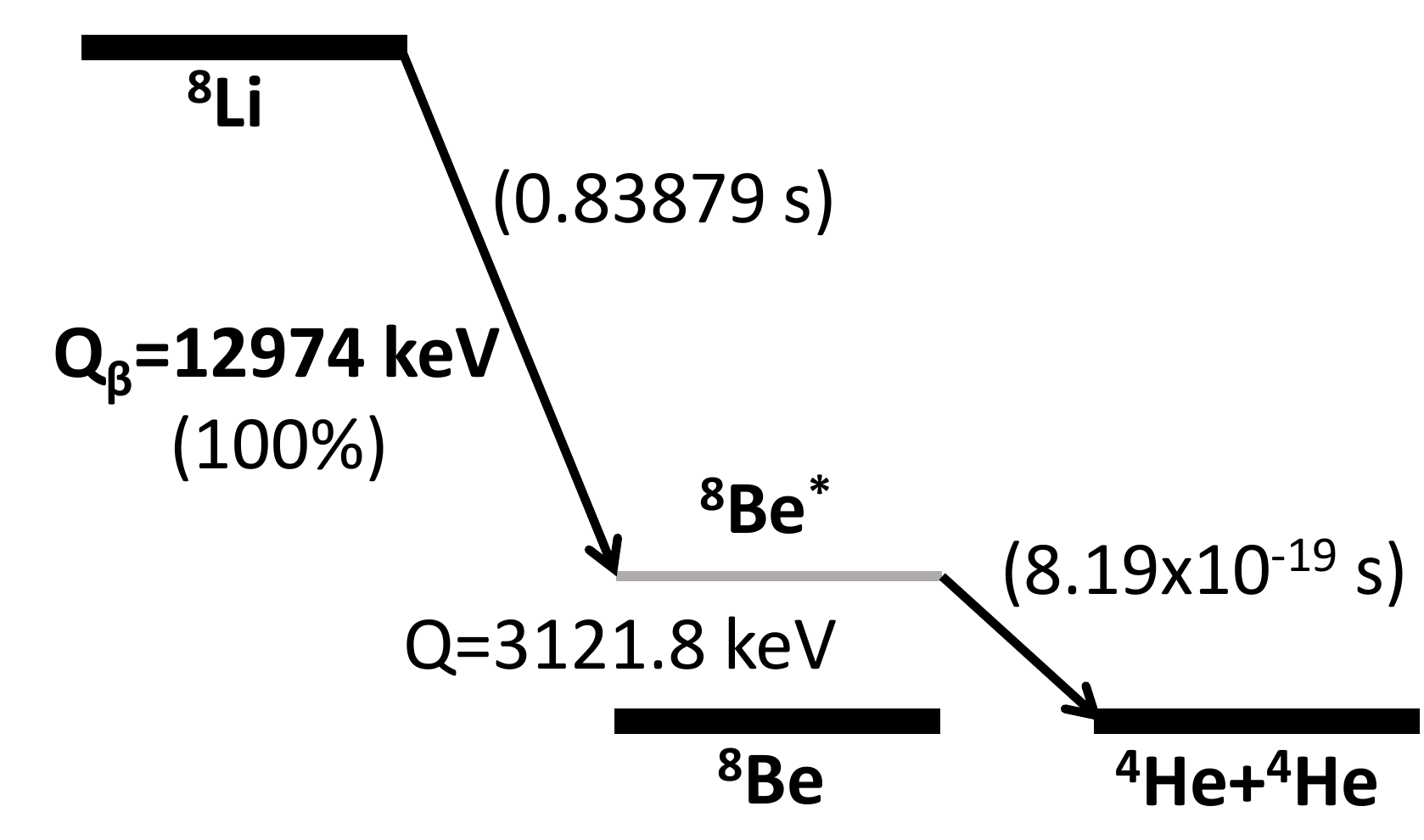} 
\end{minipage}
\caption{The decay scheme of isotopes chain $^{144}{\rm Ce}-$$^{144}{\rm Pr}$~\cite{Ce_source} and $^8{\rm Li}$ isotope.}
\label{fig_1}
\end{figure}

As a source of $\bar\nu_e$, the decay chain of isotopes $^{144}{\rm Ce}-$$^{144}{\rm Pr}$ is a suitable option for oscillometry experiment.
The decay scheme is shown on the left panel of Fig.~\ref{fig_1}.
The antineutrino energy spectrum of $\rm{^{144}Pr}$ is continuous with the end point around 3 MeV and with an overall half-life of 285 days.
About 48.5\% of the emitted antineutrinos are at energies above the detection threshold of the inverse beta decay (IBD) reaction (the value of the threshold is 1.8 MeV) and thus can be used for the measurements.
Based on the SOX calculation, the maximal source activity can reach 100 kCi.
We assume two values of activity, 50 and 100 kCi, which depends on the experimental configuration.
The detailed information about source production is presented in~\cite{Ce_source,Noto:2016juu}.

\subsubsection{$^8{\rm Li}$ source}

Lithium-8 is a short-lived nuclide with half-lifetime 0.838 s and $Q$-value 16004.13 keV. 
$^8{\rm Li}$ always decays to the exited state of $^8{\rm Be}$ with $Q_\beta= 12974$ keV~\cite{nucl_data}.
The decay scheme is depicted on the right panel of Fig.~\ref{fig_1}.
The antineutrino spectrum from $^8{\rm Li}$ beta-decay is continuous with an end point of 13 MeV~\cite{Zhao:2015bba}.
Based on the IsoDAR proposal, lithium-8 can be produced in two nuclear reactions: direct production $p+$$^8{\rm Be}\rightarrow$$^8{\rm Li}+2p$ and neutron capture by $^7{\rm Li}$.
Proton beam from a superconducting cyclotron (10 mA, 600 kW) hits beryllium target, which is surrounded by sleeve with isotropically distributed lithium-7.
We use an improved sleeve design, which is a cylinder with 1.3 m length and 1.2 m diameter~\cite{new_design}.
This design allowed us to increase the $^8{\rm Li}$ production up to 0.019 atoms per one proton. 
The sleeve is put inside a container with layers of graphite, iron and concrete.
It captures almost all escaping fast neutrons from the sleeve, more detailed information about sleeve and container design can be found~\cite{Abs:2015tbh}. 

\subsection{Jinping facility as a detector for the oscillometry experiment}
\par
\indent

The Jinping neutrino detector will be located in the Jinping Mountain, Sichuan Province, China with maximum overburden around 2400 meters~\cite{Jinping}.
It focuses on low energy neutrino and antineutrino measurements like solar, supernova and geo neutrinos.
It has the lowest muon flux $\simeq2 \times 10^{-10}~\rm{cm^{-2}\cdot s^{-1}}$ comparing to other experiment facilities.
The Jinping collaboration plans to build a 2 kton detector using slow liquid scintillator (LSc)~\cite{slow}.
This delays the scintillation process and thus separates from the Cherenkov light.
This can significantly increase the background rejection capability using the particle identification method~\cite{slow}.
The inner detector volume will have a spherical shape with a radius around 8.2 meters.
The expected energy resolution will be $5\%/\sqrt{E[{\rm MeV}]}$.
The Jinping detector will use PMTs with $\sim 1$ ns TTS.
This in turn will allow the Jinping detector to reach a position resolution of $10~{\rm cm}/\sqrt{E[{\rm MeV}]}$ (this value is used in our calculations).
In case of IsoDAR the position resolution should be enlarged due to additional part comes from initial source position uncertainty around 40 cm. 

\section{Layout of the experiment and numerical analysis}
\label{sec_3}

\par
\indent

\subsection{Experimental setup}
\par
\indent

\subsubsection{Point-like $^{144}$Ce source}
\par
\indent

\begin{figure}[ht]
\centering\includegraphics[scale=0.5]{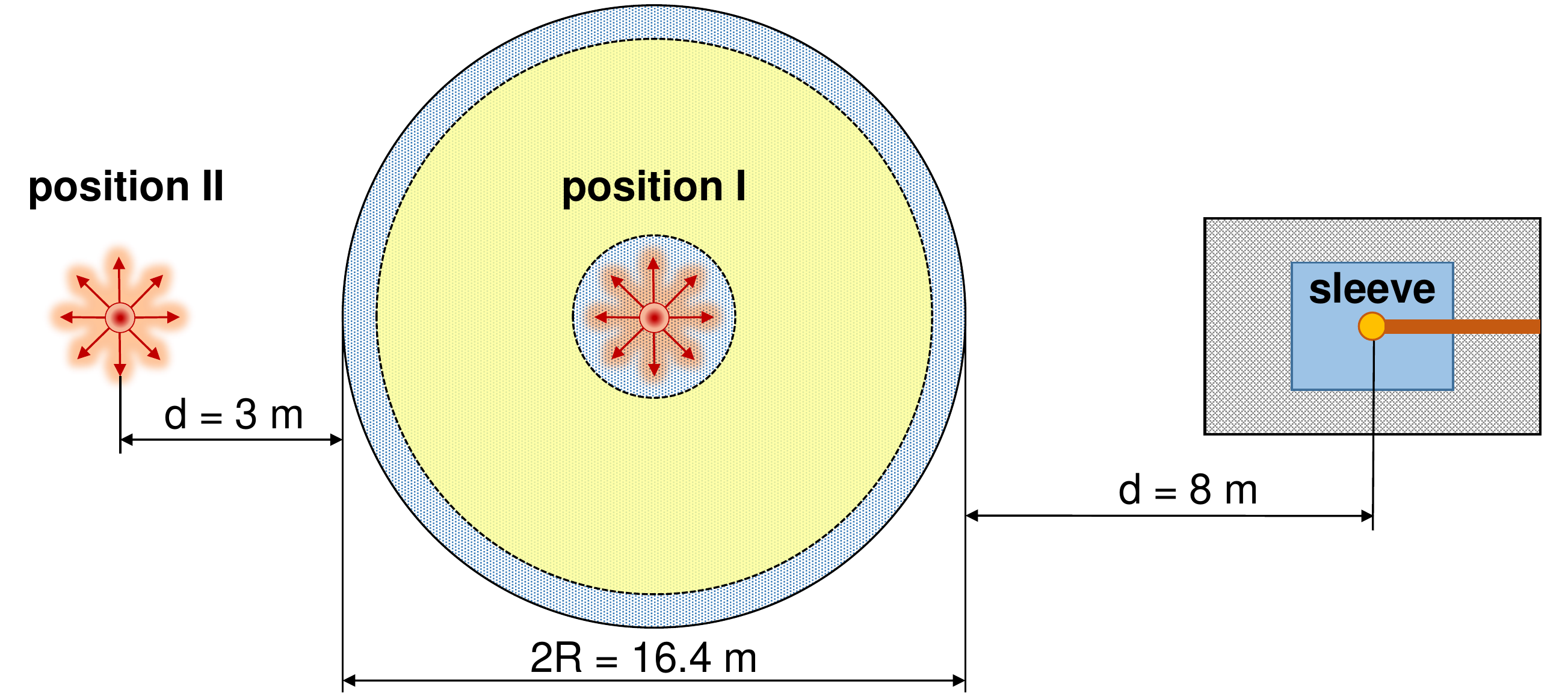}
\caption{The schematic layout of the proposed experiment. Two positions for the radioactive source are assumed: at the center (position I) and outside of the detector (position II). Two fiducial volume cuts are applied. Inner cut with radius 100 cm  and outer cut with width 70 cm. Yellow area is active volume for position I, yellow area plus inner cut region is active volume for position II. The IsoDAR container locates further outside in comparison with position II. Only outer cut is applied for this case.}
\label{fig_2}
\end{figure}

The scheme of the experiment is depicted in Fig.~\ref{fig_2}.
There are two possible locations for point source, one at the center of the sphere (position I) and another at a distance 3 m from the edge of the detector (position II).
The value of 3 meters is conservative and comparable to similar experimental proposals~\cite{Borexino:2013xxa,celand}.
The source at the detector center case gives a higher statistics but a shorter range of baseline.
Technically, the more realistic case is when the source is outside of the detector.
This configuration allows us to not break the structure of the detector inner volume.
For both source positions the exposure time is assumed to be 450 days.
Initial source activity will be 50 and 100 kCi  for position I and position II respectively.
The expected non-oscillation event rate is 28.5K and 73.8K for position I and position II respectively.


Concerning the background, as we consider the antineutrino emitter in the energy range the 1.8--3 MeV, the background will be geo-antineutrinos and antineutrinos from reactors.
One of the main advantages of Jinping is a location far from any working nuclear power plants, which decreases the reactor flux significantly.
Our estimation of the total background gives a value of $\approx100$ events per 450 days, which is two order of magnitude smaller than the expected signal rate.
Considering these things the experimental setup can be considered free from IBD background.

Besides, as the $^{144}{\rm Ce}-$$^{144}{\rm Pr}$ chain has gamma radiation and the source activity is huge, so this experimental setup can also be used for searching dark matter and dark photon as proposed in ref.~\cite{Pospelov:2017kep}.

\subsubsection{IsoDAR}
\par
\indent

Regarding the IsoDAR setup, since the container with the sleeve is much larger than a point source, there is only one option for this source is to be installed outside of the detector active volume.
The distance between beryllium target inside sleeve and the detector edge is 8 m.
This value is comparable with KamLAND plus IsoDAR setup.
The working time is assued to be 5 years with 90\% duty cycle.
Only outer fiducial volume cut for the Jinping detector is applied for the $^8{\rm Li}$ source.
The expected amount of events in 5 years is $1.76\cdot10^6$, which is in two orders of magnitude larger then for cerium source.

The $^{8}$Li production is exponentially decreasing as a function of the distance to the target center. Such a decreasing shape depends on the neutron capture cross section, which is known to a large extent.
The proton beam can produce fast neutrons that penetrating the sleeve. Those fast neutrons can be absorbed with the iron shielding surrounding the sleeve volume.
Further block can be placed downstream to the beam in order to completely remove the beam related neutron background.

\subsection{Event rate calculation with toy MC}
\par
\indent

\subsubsection{A point source  like $^{144}$Ce}
\par
\indent

Analytically the equation for expected event rate from a source with initial activity $A_0$ can be expressed as~\cite{Ce_source}:
\begin{equation}
\label{eq_3}
N(L,E)=\iint\frac{A_0}{\lambda}(1-e^{-\lambda t_e})\rho\frac{\sigma_{\rm IBD}(E)\cdot S(E)\cdot p(L,E)}{4\pi L^2}dVdE,
\end{equation}
where $L$ is distance between point source and a point inside the detector; $E$ is the antineutrino energy; $\lambda$ is a constant of radioactive decay; $t_e$ is an exposure time (450 days); $\rho$ is the density of free protons per cubic meter of LSc ($6.4\cdot10^{28}$ for LAB~\cite{juno}); $\sigma_{\rm IBD}$ is IBD cross-section~\cite{IBD_xsec}; $S(E)$ is the spectrum shape of released antineutrinos~\cite{Ce_source}, which is normalized to one; $p(L,E)$ is the oscillation probability Eq.~\eqref{eq_1}.

The event rate calculation for position I is quite simple using Eq.~\eqref{eq_3}, as it is a spherically symmetric case.
However for more a complicated geometry like position II this method  is not suitable and requires time consuming calculations.
A much simpler approach uses MC toy events.
Let's connect the coordinate system with the source position, assuming that source is a point.
Then randomly distribute events inside 4D volume for {\it i-th} event we have four coordinates $(x_i,y_i,z_i,E_i)$.
After that we count all events which are inside our spherical volume with energies in the range [1.8, 3] MeV.
Each interesting MC event has its own weight.
The weight can be easily calculated using next formula:
\begin{equation}
\label{eq_4}
\omega_i=\frac{\sigma_{\rm IBD}(E_i)\cdot p(E_i, L_i)}{4\pi(x_i^2+y_i^2+z_i^2)},
\end{equation}
where $L_i^2=x_i^2+y_i^2+z_i^2$. Using the weight definition the desired event rate can be rewritten as:
\begin{equation}
\label{eq_5}
N(L,E)={\cal C}\cdot\frac{\sum\omega_i}{N_{MC}^{in}}\int dV\int_{1.8}^{3.0}S(E)={\cal C}\cdot\frac{\sum\omega_i}{N_{MC}^{in}}\cdot\frac{4}{3}\pi R^3\int_{1.8}^{3.0}S(E),
\end{equation}
where $\cal C$ carries all constants from Eq.~\eqref{eq_3}; $N_{MC}^{in}$ is the number of MC events inside the desired 4D volume. 
The second factor effectively represents the weighted event number per volume per energy.


In this way the MC approach demonstrates independence from geometry and allows us to easily apply event by event smearing, which is associated with energy and position resolutions.
Also this allows us to use one dimensional chi-square function with only one variable $L/E$ instead of the traditional two dimensional chi-square function with two separated variables $L$  and $E$~\cite{Ce_source}.

\begin{figure}[ht]
\centering
\begin{minipage}[b]{0.44\linewidth}
\centering
\includegraphics[scale=0.41]{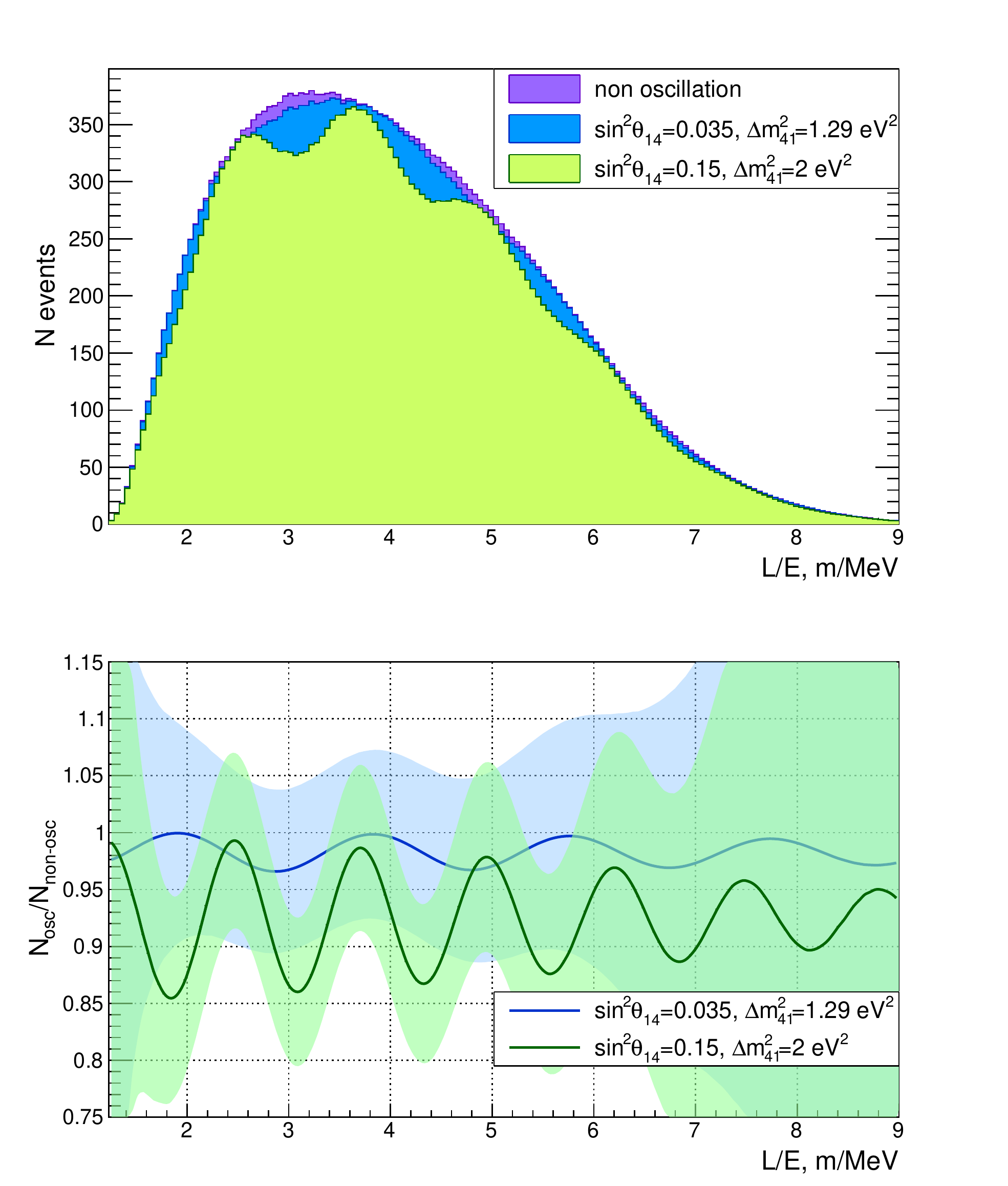}
\end{minipage}
\qquad
\begin{minipage}[b]{0.48\linewidth}
\centering
\includegraphics[scale=0.41]{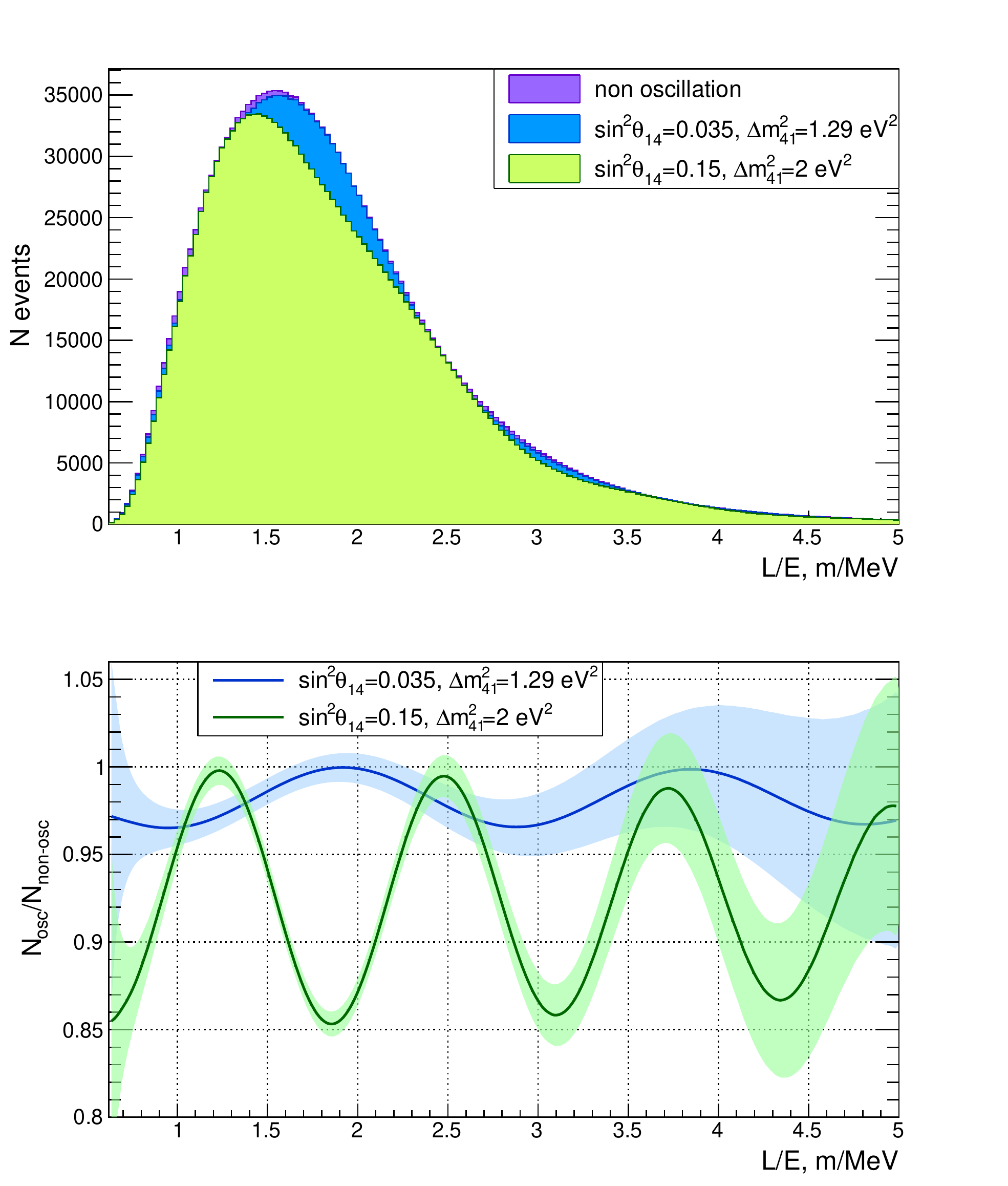} 
\end{minipage}
\caption{The top left panel corresponds to the non-oscillation event rate and two oscillation rates with the different values of oscillation parameters as a function of $L/E$ variable (the right panel is for IsoDAR case).
The blue spectrum is based on current best fit values for the sterile neutrino~\cite{Dentler:2018sju}.
The bottom left panel corresponds to the appropriated oscillation curves as a function of $L/E$ variable (the right panel is for IsoDAR case). Colored bands for the bottom pictures correspond to one sigma statistical uncertainty.}
\label{fig_3}
\end{figure}

The expected event rate for the position II and appropriated oscillation curves are depicted in the left part of Fig.~\ref{fig_3} and the right part corresponds to the lithium source.
A second pair of oscillation parameters is an example, which demonstrates larger oscillations.
As can be seen from the bottom Fig.~\ref{fig_3} smearing decreases the oscillation amplitude with growth of $L/E$ variable.

\subsubsection{IsoDAR}
\par
\indent

As the IsoDAR is not a point-like source anymore, the baseline of each event depends on the location of the isotope.
In Eq.~\eqref{eq_3} the term with activity should be substituted by the total number of antineutrinos produced for 5 years inside the sleeve volume.
For the MC calculations all initial events should be randomly distributed inside the sleeve volume in the energy range [1.8; 12.9] MeV.
We assume an exponential distribution of lithium-8 atoms density  inside sleeve from the beryllium target to the sleeve's edge.
Then the weight of $i$-th event can be written as:
\begin{equation}
\label{eq_6}
\omega_i^s=\frac{\sigma_{\rm IBD}(E_i)\cdot p(E_i, L_{di})\cdot\rho_{8Li}(x_i,y_i,z_i)}{4\pi((x_d-x_i)^2+(y_d-y_i)^2+(z_d-z_i)^2)},
\end{equation}
where $s$ stands for sleeve case; $\rho_{8Li}$ -- exponential density of lithium atoms;  
$x_d,y_d,z_d$ -- coordinates of a point inside the detector; $L_{di}$ -- distance between points inside the sleeve and inside the detector.
After that the Eq.~\eqref{eq_5} can be rewritten as:
\begin{equation}
\label{eq_5}
N(L,E)={\cal C}\cdot\frac{\sum\omega_i^s}{N_{MC}^{in}}\cdot\frac{4}{3}\pi R^3\cdot V_s\int_{1.8}^{12.9}S(E),
\end{equation}
where $V_s$ -- sleeve volume.
As the IsoDAR case has much larger statistics, it allows us to measure oscillation parameters for sterile neutrinos with better precision than that for the radioactive source.

\subsection{Statistical evaluation}
\par
\indent

For the sensitivity definition to sterile neutrino parameters we follow the so-called classical method of calculating a confidence level. 
This method is based on the calculation of a $\Delta\chi^2$ function, which, as Wilks's theorem predicts~\cite{wilks}, should follow a chi-square distribution.
The number of degrees of freedom of the $\Delta\chi^2$ distribution can be calculated as the difference between degrees of freedom (dof) of initial chi-square functions.
In the case of sterile neutrinos there are only two parameters $\theta_{14}$ and $\Delta m^2_{41}$, so dof equals to 2.
It is worth noting that the conditions required by Wilks's theorem are usually not fulfilled in a sterile neutrino search. 
Instead of the above mentioned method, a conventional way to set exclusion limits is to use Feldman-Cousins (FC) procedure, as described in~\cite{Feldman-Cousins}. 
We have carried out MC simulations for  $\Delta\chi^2$ using FC procedure and figured out that it approximately follows an analytical distribution of chi-square function with 2 dof.   
Thus it is justified to use an analytical assumption instead of the time consuming FC approach in our calculations. 
The correspondence between values of $\Delta\chi^2$ and a significance level was taken from the Table 39.2 in~\cite{PDG}.

Using an approach similar to that used in~\cite{Ce_source}, we define one dimensional chi-square function in the next form:
\begin{equation}
\label{eq_6}
\chi^2=\sum\frac{(N_i^{\rm obs}-N_i^{\rm pre})^2}{N_i^{\rm pre}(1+N_i^{\rm pre}\cdot\sigma_b^2)}+\frac{\alpha^2}{\sigma_\alpha^2},\qquad N_i^{\rm pre}=(1+\alpha)S_i,
\end{equation}
where $N_i^{\rm obs}$ -- number of events with non-oscillation assumption; $S_i$ -- number of events with oscillation assumption; $\sigma_b$ -- uncorrelated bin to bin uncertainty, which includes all possible shape fluctuations. In our case we use 2\% as value for $\sigma_b$ (0\% for IsoDAR case); $\alpha$ -- so-called nuisance parameter associated with the source activity;  $\sigma_\alpha$ -- systematic error for the source activity (2\% and 5\% for radioactive source and IsoDAR respectively).
The choice of a quite large value of $\sigma_b$ for the radioactive source is used to overlap all possible systematic effects, which can be associated with energy scale (non-linearity of the energy response, which is typical for LSc detectors), spill in or out effect (leakage of IBD events near detector edge) etc.

In general the sensitivity to sterile neutrinos can be calculated through the minimization of a $\Delta\chi^2$ function, which is given by:
\begin{equation}
\label{eq_7}
\Delta\chi^2=\chi_{\rm fix}^2-\chi_{\rm min}^2,
\end{equation}
where  $\chi_{\rm fix}^2$ uses a fixed pair of oscillation parameters; $\chi_{\rm min}^2$ uses best fit values for oscillation parameters.
Both chi-square functions are identical to Eq.~\eqref{eq_6} and should be minimized through all nuisance parameters.
In this analysis we use the so-called Azimov data set, consequently $\chi_{\rm min}^2$ equals to zero and $\chi_{\rm fix}^2$ defines the sensitivity completely.

\subsection{Systematic uncertainties treatment}
\par
\indent

Neutrino oscillation experiments mostly concern three systematic uncertainty sources: neutrino flux, neutrino interaction and the detection process.
In the source-based sterile neutrino search in a LSc detector, as the neutrino IBD interaction cross-section is known precisely,
the systematic uncertainties come from the other two main sources.
They are classified into three categories and described as below.

The first one is the flux, which relates to the knowledge on the source neutrino emission precision. 
We assume a uniform 1.5\% systematic uncertainty on the knowledge of the neutrino flux for the cerium source. 
We expect an increased knowledge of the neutrino emission from the radioactive source at the time that the new generation of low-background neutrino experiments such as Jinping starts data taking. 
At the time being now, a 1.5\% systematic uncertainty is adopted by various
sensitivity studies performed for different source neutrino projects.
A 5\% systematic uncertainty is assumed for IsoDAR. 
It is based on the previous studies of this topic referring to~\cite{Conrad:2013ova}.

The second is the detection efficiency uncertainty. 
We assume a 90\% efficiency across the energy range for all neutrino detection with
a systematic uncertainty along the full energy range for both sources. 
Usually this systematic uncertainty is at one percent level.
Since this systematic error is fully correlated with flux uncertainty, we decided not to double this. 
The $\sigma_{\alpha}$ accounts for this effect.
This detection efficiency includes both prompt positron and delayed neutron detection. 
A spill in/out effect can happen when the interaction vertex is inside a volume while the neutron capture is outside the fiducial volume. 
We examined the spill effect and it turns out to be negligible. As the neutron moves with a distance typically less than a few centimeters, the spill effect only takes place in the boundary shell with 20-ish cm.
With the radioactive source at a distance 2.3 m away from the detector surface, the overall spill-in effect gives 3\% more events at the outer 20 cm shell. 
In the case of the source located in the center of the spherical detector, we will need a fiducial volume cut for the inner sphere around the source, the spill in effect occurring here provides us 19\% more events for the first inner 20 cm shell. 
These spill-in effects have been included and tested. 
There was no noticeable impact from them considering a 100\% uncertainty on those spill-in events.

The last one is the energy scale and resolution uncertainties. 
The energy resolution is based on the Jinping proposal,
which is 5\% at 1 MeV. 
We tested the sensitivity with 5\%, 8\% and 10\% resolutions. 
That result guides us to ignore the resolution systematic uncertainty for this case.
Regarding the energy scale uncertainty, 1\% uncertainty is assumed based on the Daya Bay results~\cite{Adey:2019zfo}. 
Two independent approaches have been performed to cross
check the energy scale uncertainty. 
The first of them is the shifting of the reconstructed neutrino energy event by event with $1\sigma$ uncertainty of 1\%. 
This might be computing intensive because of the fact that in each fitting iteration, the energy scale parameter changes and all events should be updated based on the current energy scale parameter.
The second is to apply such an energy scale uncertainty as a 2\% bin-to-bin uncorrelated error ($\sigma_b$). 
The results from both approaches are consistent. 
Therefore, the results shown in this paper have been obtained using the second approach due to the reasons of simplicity and calculation speed.
It should be mentioned that lithium source shape uncertainties are much smaller, so there are no reasons to use this $\sigma_b$ together with IsoDAR.

\section{Results and discussion}
\label{sec_4}
\par
\indent

Following the statistical method mentioned above, the general sensitivity to sterile neutrinos can be represented as a two-dimensional exclusion plot.
\begin{figure}[ht]
\centering
\begin{minipage}[b]{0.45\linewidth}
\centering
\includegraphics[scale=0.41]{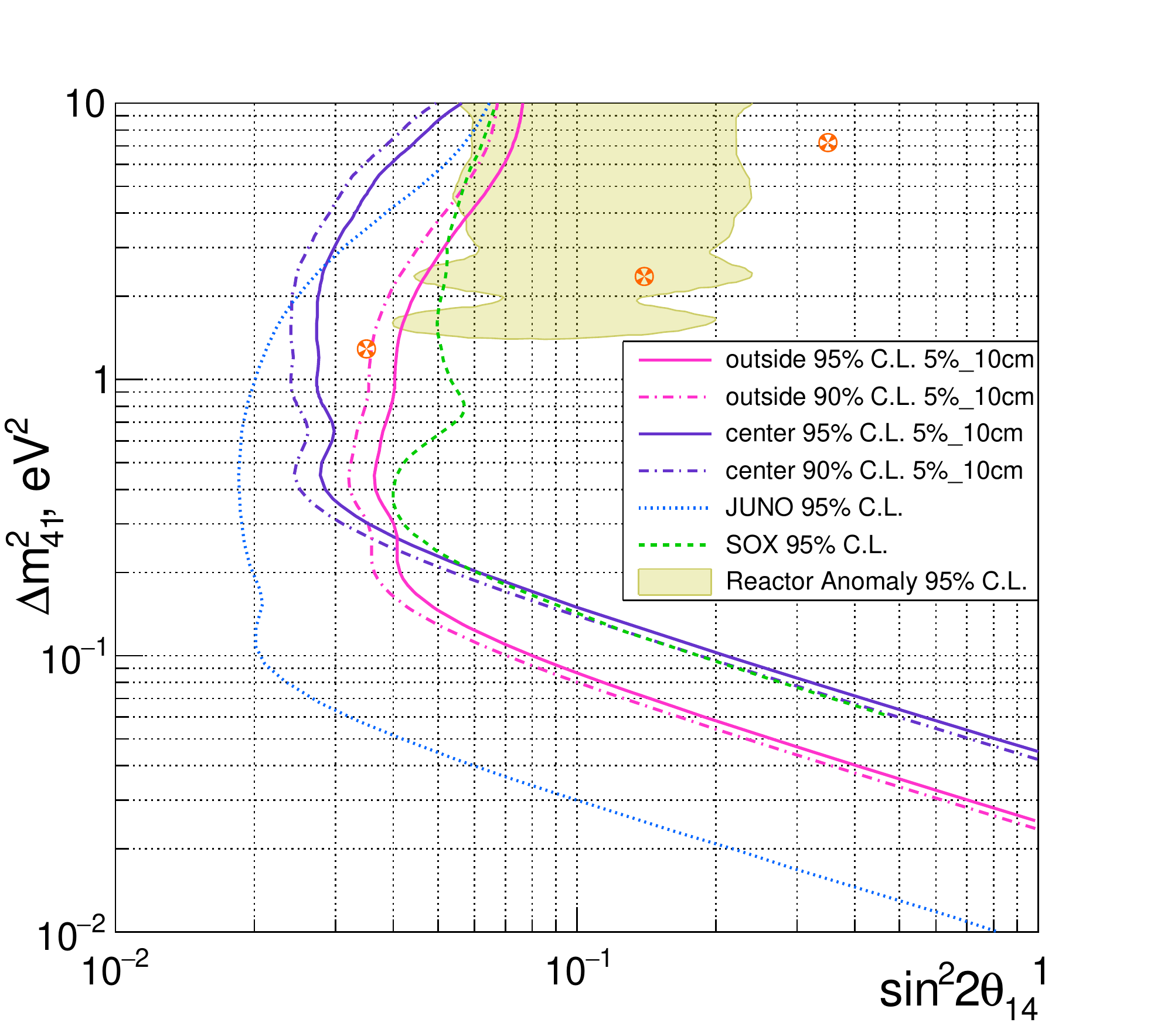}
\end{minipage}
\qquad
\begin{minipage}[b]{0.49\linewidth}
\centering
\includegraphics[scale=0.41]{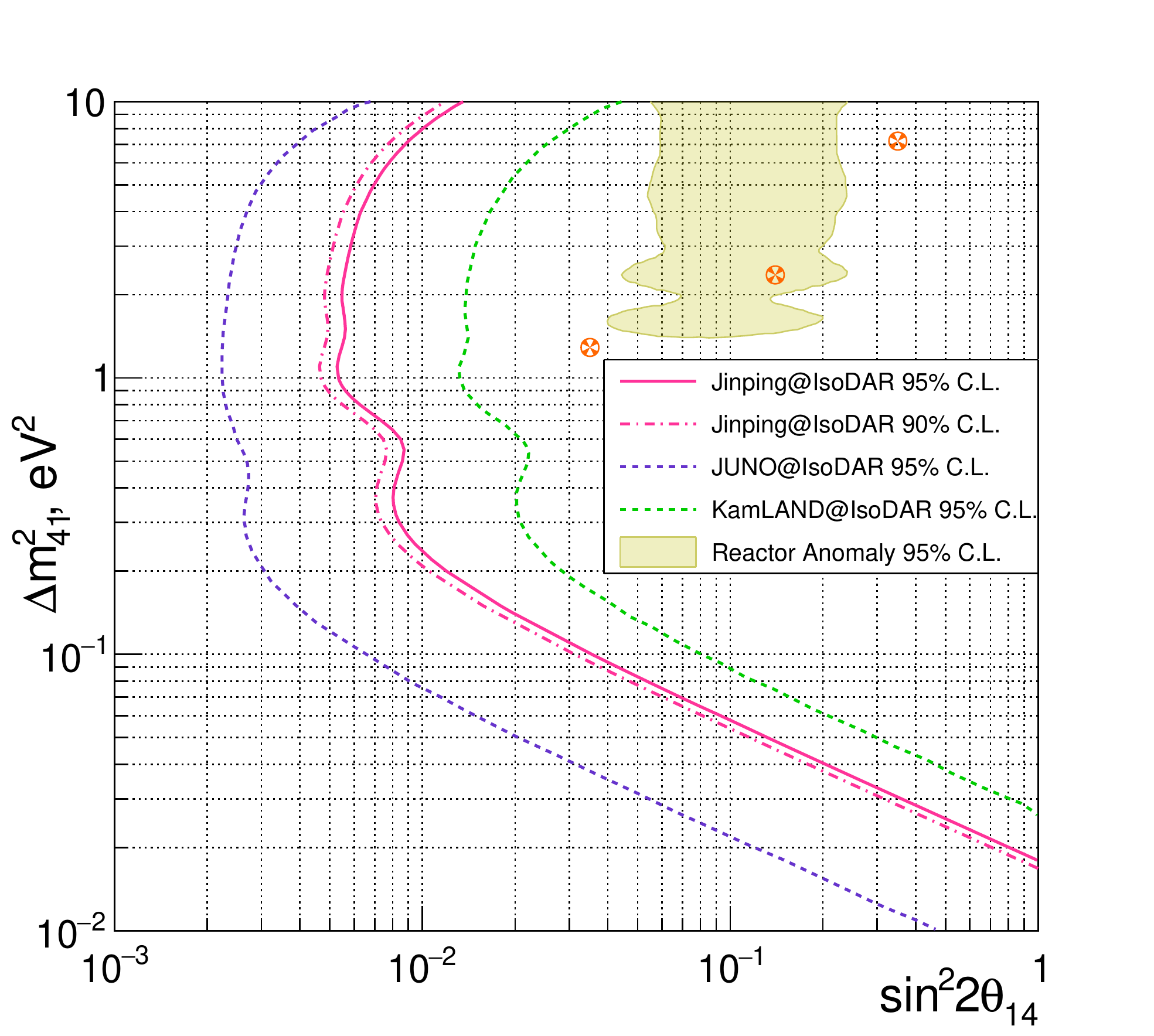} 
\end{minipage}
\caption{The exclusion contours on a two-dimensional parameter space. 90\% and 95\% C.L. are shown for two possible setups. The left panel corresponds to the radioactive source and the right part corresponds to the IsoDAR. Combined reactor anomaly is also shown~\cite{re_anom}. The stars indicate the current best fit value of the sterile neutrino~\cite{Dentler:2018sju}, the best value of the reactor anomaly~\cite{re_anom} and the best fit result of the Neutrino-4 experiment~\cite{neutrino4}, from left to right respectively.}
\label{fig_4}
\end{figure}
For comparison, the SOX and JUNO results are also drawn on the left panel of Fig.~\ref{fig_4}, the KamLAND and JUNO plus IsoDAR results are drawn on the right panel of Fig.~\ref{fig_4}. 
As can be seen, all currently interesting regions for sterile neutrino searching are covered if we consider position I.
For position II, some part of the reactor anomaly and best fit are still not fully covered.
As expected, the IsoDAR setup has better sensitivity and can investigate wider ranges of the oscillation parameter phase space than the radioactive source.
Both setups can validate or reject the measured result from the Neutrino-4 experiment~\cite{neutrino4}.
To avoid any problems associated with heterogeneity of position and energy resolution inside the detector, we apply a fiducial volume cut: outer cut 70 cm for both positions of the radioactive source and IsoDAR, inner cut 100 cm only for position I for the radioactive source.

\begin{figure}[ht]
\centering\includegraphics[scale=0.76]{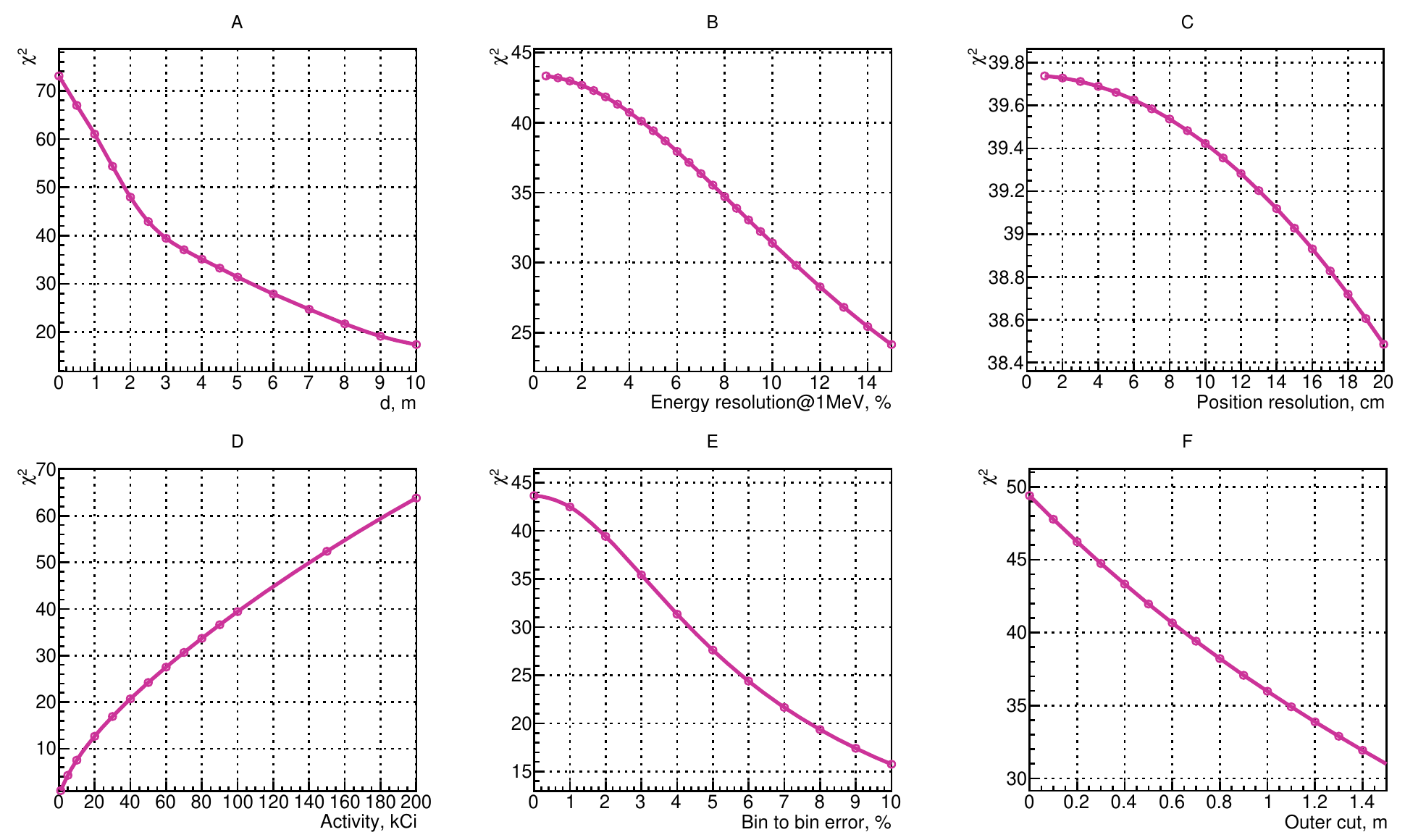}
\caption{The significance of the experiment as a function of different parameters. Panel A corresponds to the distance between the source and the detector edge. Panel B corresponds to the energy resolution of Jinping. Panel C corresponds to the position resolution of Jinping. Panel D corresponds to the source activity. Panel E corresponds to the value of bin to bin uncertainty. Panel F corresponds to the length of outer cut. Assumed oscillation parameters  $\sin^2(2\theta_{14})=0.1$, $\Delta m^2_{41}=1~{\rm eV}^2$.}
\label{fig_5}
\end{figure}

In addition, we investigated the significance dependency as a function of different parameters such as energy resolution, position resolution, distance between source and the detector edge for position II, source activity, bin to bin uncertainty and the length of outer cut. 
All results are summarized in Fig.~\ref{fig_5}.
The source should be located as close as possible to the detector edge for position II.
Energy resolution is an important parameter for this analysis, as larger value of the energy resolution erodes the spectral shape and causes the loss of information.
Spectrum shape should be known precisely.
And position resolution of the detector should not be excellent for oscillometry measurements.
All these conclusions are valid for setup with the IsoDAR lithium source.

\section{Conclusion}
\par
\indent

Our research has demonstrated the success of using a powerful antineutrino radioactive source and the IsoDAR source for the sterile neutrino search.
It is important to measure the first few units of L/E, because the waves of oscillation curve fade out very quickly due to the smearing.
The method of neutrino oscillometry allows us to observe a whole oscillation L/E spectrum inside the detector with the sterile neutrino hypothesis.
A high-intensity source and a large detector are needed to limit the statistical fluctuation and certain knowledge of the experiment performance is needed in order to control the systematic uncertinaties.
In comparison to the reactor experiments, this method is free from the large uncertainty of the reactor spectrum prediction. 

We have shown that the combination of a hand-made source and the Jinping facility can potentially provide a remarkable coverage of the sterile neutrino oscillation parameter phase space hence an exploration on the previous sterile neutrino anomaly results.
Besides, this setup will provide us a clear test of the Neutrino-4 results.
Therefore, it may be a considerable part of the Jinping's scientific programs.
The implementation of such an experiment may also deliver a conclusive result to the light sterile neutrino search.

\acknowledgments

This work was supported by the National Key R\&D program of China under Grant No. 2018YFA0404103, the National Natural Science Foundation of China under Grant No. 11775315, Key Lab of Particle \& Radiation Imaging, Ministry of Education under project No. 20190105. We thank to the State University of the New York research foundation and DOE, USA. We would like to express our greetings to Prof. Jiajun Liao for the important suggestions and discussions in improving the paper.

\end{document}